\documentclass[12pt]{article}%
\usepackage[utf8]{inputenc}
\usepackage{amsmath}
\usepackage{amssymb}
\usepackage{geometry}
\usepackage{color}
\usepackage[doublespacing]{setspace}
\usepackage{amsfonts}
\usepackage{graphicx}%
\providecommand{\U}[1]{\protect\rule{.1in}{.1in}}
\makeatletter

\newtheorem{corollary}{Corollary}

\newtheorem{example}{Example}

\newtheorem{lemma}{Lemma}

\newtheorem{proposition}{Proposition}

\newtheorem{remark}{Remark}

\makeatother
\makeatother
\makeatother
\begin{document}

\title{P-Bubbles, Q-Bubbles, and Risk Premia}
\author{Robert A. Jarrow
\ \ \ \ \ \ \ \ \ \ \ \ \ \ \ \ \ \ \ \ \ \ \ \ \ \ \ \ \ \ \ \ Simon S. Kwok\\\ \ \ \ \ Cornell
University\ \ \ \ \ \ \ \ \ \ \ \ \ \ \ \ \ \ \ \ \ \ \ \ The University of
Sydney\thanks{We thank Jiti Gao, Peter Phillips, Roberto Ren\`{o}, Shuping
Shi, Jun Yu, and all the seminar and conference participants. Their
constructive comments led to substantial improvement of the paper. Jarrow:
Samuel Curtis Johnson Graduate School of Management, Cornell University,
Ithaca, N.Y. 14853; Email: raj15@cornell.edu. \ Kwok: School of Economics, The
University of Sydney, NSW 2006, Australia; Email: simon.kwok@sydney.edu.au.}}
\date{First version: February 2025\\
This version: April 2026}
\maketitle

\begin{abstract}
We develop a unified modeling framework that connects two distinct types of
bubbles defined in the literature: the rational bubbles (\emph{aka}
$\mathbb{P}$-bubbles), and the local martingale bubbles (\emph{aka}
$\mathbb{Q}$-bubbles). We show that the local martingale bubble model includes
the classical rational bubble as a special case. We relate both types of
bubbles to an equity's risk premium via a novel decomposition.\medskip{}

\bigskip{} \bigskip{}


\bigskip{} \pagebreak{}

\end{abstract}

\section{Introduction}

There is a substantial literature that models stock price bubbles. Due to the
latent nature of stock price bubbles, different definitions\footnote{See
Baumann and Janischewski (2025) for a recent survey on the definitions of
bubbles.} have been explored and different methodologies are used for making
inferences about bubbles. Two bubble theories stand out. The first is the
rational bubble theory. In an infinite horizon model, this theory generates
asset price bubbles as a component within the present value of future cash
flows where, due to the transversality condition, the liquidation value of
asset at infinity is strictly positive. The existence of such bubbles
(\emph{aka} $\mathbb{P}$-bubbles hereafter) is consistent with the standard
no-arbitrage condition (e.g., Blanchard (1979), Flood and Garber (1980),
Shiller (1981), Blanchard and Watson (1982)). Bubble detection is possible
using historical asset prices under the statistical probability measure
$\mathbb{P}$ (e.g., West (1987), Diba and Grossman (1988), Phillip, Wu and Yu (2011)).

The second is the local martingale theory of bubbles. This theory is developed
in a continuous time model (e.g., Lowenstein and Willard (2000), Cox and
Hobson (2005), Heston, Lowenstein and Willard (2007), Jarrow, Protter and
Shimbo (2010); see Protter (2013) for a review). Assuming a stronger
no-arbitrage condition known as no-free-lunch-with-vanishing-risk (NFLVR), the
first fundamental theorem of asset pricing (Delbaen and Schachermayer (1994))
guarantees the existence of an equivalent local martingale measure
$\mathbb{Q}$ in which the asset price process (plus dividends, discounted at
the risk-free rate) is a local martingale. A bubble (\emph{aka} $\mathbb{Q}%
$-bubbles hereafter) arises when the asset price process is a strict local
martingale (or equivalently a strict supermartingale if price is nonnegative),
as distinct from an ordinary martingale, under $\mathbb{Q}$. Inference
procedures for $\mathbb{Q}$-bubbles have been developed (e.g., Jarrow and Kwok
(2021), Fusari, Jarrow and Lamichhaine (2024)).

The purpose of this paper is to develop a unified modeling framework for
simultaneously analyzing both types of bubbles. Using a continuous-time
setting and assuming NFLVR, we show that the equity's price can be decomposed
into the sum of three distinct components: the fundamental value (defined as
the expected discounted future dividends under $\mathbb{P}$), a $\mathbb{P}%
$-bubble, and a $\mathbb{Q}$-bubble. Herein, the classical rational theory
bubble is obtained as the special case where the model's horizon approaches
infinity and the $\mathbb{Q}$-bubble is set to zero.

The new framework clarifies the roles played by both types of bubbles in
characterizing the equity risk premium (ERP), defined as the difference
between the equity's expected return under $\mathbb{P}$ and under $\mathbb{Q}%
$. The ERP can be decomposed into two distinct components: (1) the
$\mathbb{P}$-component, which contains both the expected return on the
fundamental value and the expected increment in the $\mathbb{P}$-bubble; and
(2) the $\mathbb{Q}$-component, which captures the expected decrease in the
discounted $\mathbb{Q}$-bubble. Both the $\mathbb{P}$- and $\mathbb{Q}%
$-bubbles contribute to the required compensation for price risk under the
local martingale bubble theory. We show that the ERP obtained in the local
martingale bubble theory is generally greater than the ERP derived in the
rational bubble theory, which rules out $\mathbb{Q}$-bubbles. This ERP
characterization is quite general and fully nonparametric.\footnote{Jarrow and
Kwok (2024) study the price explosiveness in $\mathbb{P}$ when $\mathbb{Q}%
$-bubbles are present. Their model for the price process under $\mathbb{P}$
assumes that the instantaneous drift rate (which determines the equity risk
premium) is linear in the price process' quadratic variation.}

Our work complements to the literature on ERP estimation using market data.
There have been proposals to use option price data instead of historical and
accounting data to compute ERP. For example, Martin (2017) obtains a lower
bound for ERP based on the risk-neutral variance of return (a second moment
property) under the standard no-arbitrage condition. Our analysis shows that,
under the stronger NFLVR condition, this lower bound needs to be revised
upward by an amount that reflects the negative risk-neutral expected return (a
first moment property) due to the presence of a $\mathbb{Q}$-bubble.

An outline for this paper is as follows. Section 2 describes the rational
bubble model and introduces the $\mathbb{P}$-martingale deviation which is
useful for measuring the expected change in P-bubble. The model is first
studied in discrete time with a constant discount factor (\emph{aka} the
classical rational bubble model; Section 2.1-2.2). It is then generalized to
the continuous-time setting with a stochastic discount factor (Section 2.3).
Section 3 presents the local martingale bubble model, which is shown to
include the rational bubble model as a special case.\ Section 4 characterizes
the ERP in terms of the two types of bubbles. Section 5 concludes. All
technical proofs are given in the Appendix.

\section{The Rational Bubble Model}

There is a long-standing literature on rational bubbles (e.g., Blanchard
(1979), Flood and Garber (1980), Shiller (1981), Blanchard and Watson (1982),
West (1987), Diba and Grossman (1988), Santos and Woodford (1997)) using an
infinite-horizon model for a recent review, see, e.g., Miao (2014), Martin and
Ventura (2018), and Hirano and Toda (2024b). 

We first describe the model in a discrete-time setting with a constant
expected discount rate, following the classical work of Shiller (1981) and
Blanchard and Watson (1982), and then study the dynamics of the asset price
components when rational bubble exists (Section 2.1). We then introduce the
$\mathbb{P}$-martingale deviation which is related to some of the common
bubble detection methods (Section 2.2). After, we extend the discrete-time
results to the more general continuous-time framework with a stochastic
discount factor (Section 2.3). Importantly, when evaluating bubbles under the
statistical measure, we need to include a risk adjustment term, which is
easily neglected when discounting is done using an expected rate in discrete
time. In addition, the continuous-time extension is a crucial step for
constructing a unified model framework that accommodates the local martingale
bubbles (to be discussed in Section 3).

Let $T$ denote the model's horizon, which can be finite or unbounded. We are
given a filtered probability space $(\Omega,\mathcal{F}_{T},\mathbb{F}%
,\mathbb{P})$, where $\Omega$ is the state space, $\mathcal{F}_{T}$ is a
$\sigma$-algebra representing the set of events, $\mathbb{F}:=\{\mathcal{F}%
_{t}\}_{t\in\lbrack0,T]}$ is the filtration, and $\mathbb{P}$ is the
statistical probability measure.

The market has a risky asset, called the stock, which pays out dividends, and
a riskless asset. Let $S_{t}\geq0$ denote the time-$t$ ex-dividend price of
the stock. We assume that the ex-dividend stock price process $S:=\{S_{t}%
\}_{t\in\lbrack0,T]}$ is an $\mathbb{F}$-adapted process.

Let $R_{t:t+\tau}^{f}$ denote the gross risk-free rate over the period
$(t,t+\tau]$ on the riskless asset, where $\tau\leq T-t$. It is assumed to be
$\mathcal{F}_{t}$-measurable. Let $r_{t:t+\tau}^{f}:=R_{t:t+\tau}^{f}-1$
denote the associated net risk-free rate.

For $s<t$, we denote by $D_{s:t}$ the accumulated dividend value at time $t$
from re-investing the dividends received between time $s$ to $t$ at the
risk-free rate. The following relation holds true for any $s<t<u$:
\begin{equation}
D_{s:u}=D_{s:t}R_{t:u}^{f}+D_{t:u}. \label{D}%
\end{equation}

\subsection{The Classical Model in Discrete-Time}

We allow the model's horizon to be unbounded, i.e., $T\rightarrow\infty$. For
now, we consider a discrete-time setting: $t\in\{0,1,2,...\}$.\footnote{In
Section 2.3, we express the rational bubble model in continuous time with a
stochastic discount factor. As a result, an extra risk-adjustment factor needs
to be included in the key expressions.} The stock is assumed to pay out
dividends $x_{j}$ at discrete times $j=1,2,...$, where $x_{j}$ is
$\mathcal{F}_{j}$-measurable for all $j$.

The price process $S$ satisfies the difference equation:
\begin{equation}
(1+\mu)S_{t}=E_{t}^{\mathbb{P}}(S_{t+1}+x_{t+1}%
),\label{eq: recursive relation}%
\end{equation}
where $\mu>-1$ is the constant expected rate of return on stock plus
dividend.\footnote{See Blanchard and Watson (1982), equation (11.1), which
defines the constant expected rate of return on the stock plus dividend. While
setting a constant $\mu$ is a strong restriction, one can extend $\mu$ to be
stochastic and time-varying; see Section 2.3.} The conditional expectation is
taken under the statistical probability measure $\mathbb{P}$.

One solution to this difference equation is given by $S=F^{\mathbb{P}%
}:=\{F_{t}^{\mathbb{P}}\}_{t\in\lbrack0,T]}$. The quantity $F_{t}^{\mathbb{P}%
}$ represents the fundamental value of the stock at time $t$, given by
\begin{equation}
F_{t}^{\mathbb{P}}=\lim_{T\rightarrow\infty}F_{t}^{\mathbb{P}}(T)\text{,
where\ }F_{t}^{\mathbb{P}}(T):=\sum_{j=1}^{T-t}\frac{1}{(1+\mu)^{j}}%
E_{t}^{\mathbb{P}}(x_{t+j}). \label{FP}%
\end{equation}

There is a more general solution (Blanchard and Watson (1982)) given by
\begin{equation}
S=F^{\mathbb{P}}+B^{\mathbb{P}},\label{Pdecomp}%
\end{equation}
where $B^{\mathbb{P}}:=\{B_{t}^{\mathbb{P}}\}_{t\in\lbrack0,T]}$ is a
(possibly non-zero) process satisfying the recursive relation\footnote{See
Blanchard and Watson (1982), equation (11.3), or Diba and Grossman (1988),
equation (4). See also Section 2 of Phillips, Wu and Yu (2011).}:
\begin{equation}
E_{t}^{\mathbb{P}}(B_{t+1}^{\mathbb{P}})=(1+\mu)B_{t}^{\mathbb{P}}.\label{EB}%
\end{equation}
The term $B_{t}^{\mathbb{P}}$ represents the rational stock price bubble at
time $t$, defined by
\begin{equation}
B_{t}^{\mathbb{P}}=\lim_{T\rightarrow\infty}B_{t}^{\mathbb{P}}(T)\text{,
where\ }B_{t}^{\mathbb{P}}(T):=\frac{1}{(1+\mu)^{T-t}}E_{t}^{\mathbb{P}}%
(S_{T}).\label{BP}%
\end{equation}
We call $B^{\mathbb{P}}$ the $\mathbb{P}$-bubble process. The stock price
decomposition (\ref{Pdecomp}) is well defined in the sense that both limits
$F_{t}^{\mathbb{P}}$ and $B_{t}^{\mathbb{P}}$ as defined respectively in
(\ref{FP}) and (\ref{BP}) exist by the monotone convergence theorem (see lemma
\ref{lemmaA_FPBP}) below).

The time-$t$ fundamental value can be represented as the present value of the
fundamental value at a future time plus the present value of all intermittent
dividends paid from time $t$ till the future time. This is stated in the
following lemma. For notational simplicity, let $\tilde{D}_{t:t+\tau}%
:=\sum_{j=1}^{\tau}(1+\mu)^{\tau-j}x_{t+j}$ denote the total value at time
$t+\tau$ of all intermittent dividends during $(t,t+\tau]$ accumulated at a
rate of $\mu$.

\begin{lemma}
\label{lemma_FP}$F_{t}^{\mathbb{P}}=\frac{1}{(1+\mu)^{\tau}}E_{t}^{\mathbb{P}%
}(F_{t+\tau}^{\mathbb{P}}+\tilde{D}_{t:t+\tau}).$
\end{lemma}

Note that $F^{\mathbb{P}}$ may not be a $\mathbb{P}$-martingale. The dynamic
properties of $F^{\mathbb{P}}$ depend on how dividends are distributed over
time. A necessary and sufficient condition for $F^{\mathbb{P}}$ to be a
$\mathbb{P}$-martingale is provided in the following proposition.

\begin{proposition}
\label{prop_FP}(i) We have
\begin{equation}
E_{t}^{\mathbb{P}}(F_{t+\tau}^{\mathbb{P}})-F_{t}^{\mathbb{P}}=[(1+\mu)^{\tau
}-1]F_{t}^{\mathbb{P}}-E_{t}^{\mathbb{P}}(\tilde{D}_{t:t+\tau}); \label{EFF}%
\end{equation}
in particular, $F^{\mathbb{P}}$ is a $\mathbb{P}$-martingale iff
\begin{equation}
\lbrack(1+\mu)^{\tau}-1]F_{t}^{\mathbb{P}}=E_{t}^{\mathbb{P}}(\tilde
{D}_{t:t+\tau}) \label{FD_relation}%
\end{equation}
for all $\tau>0$.

(ii) If the dividend process $x$ is a $\mathbb{P}$-martingale (i.e.,
$E_{t}^{\mathbb{P}}(x_{t+j})=x_{t}$ for all $j>0$) and $x_{t}=\mu
F_{t}^{\mathbb{P}}$ for all $t$, then $F^{\mathbb{P}}$ is a $\mathbb{P}$-martingale.
\end{proposition}

\begin{remark}
\label{rmrk_FP}(i) The dynamics of $F^{\mathbb{P}}$ are a function of
$\{x_{t}\}_{t=1}^{T}$ and can be arbitrary; in particular, $F^{\mathbb{P}}$
can be a submartingale, supermartingale, or a martingale under $\mathbb{P}$.
Suppose there are no dividends over $(t,t+\tau]$ but dividends thereafter, and
$\mu\geq0$. Then $E_{t}^{\mathbb{P}}(\tilde{D}_{t:t+\tau})=0$, and
$E_{t}^{\mathbb{P}}(F_{t+\tau}^{\mathbb{P}})-F_{t}^{\mathbb{P}}=[(1+\mu
)^{\tau}-1]F_{t}^{\mathbb{P}}\geq0$, so that $F^{\mathbb{P}}$ is a
submartingale under $\mathbb{P}$. Alternatively, suppose all dividends are
paid over $(t,t+\tau]$ but no dividends thereafter. Then $F_{t}^{\mathbb{P}%
}>0$, $F_{t+\tau}^{\mathbb{P}}=0$, $E_{t}^{\mathbb{P}}(\tilde{D}_{t:t+\tau
})=(1+\mu)^{\tau}F_{t}^{\mathbb{P}}>0$, and $E_{t}^{\mathbb{P}}(F_{t+\tau
}^{\mathbb{P}})-F_{t}^{\mathbb{P}}=-F_{t}^{\mathbb{P}}<0$, so that
$F^{\mathbb{P}}$ is a supermartingale under $\mathbb{P}$.

(ii) Condition (\ref{FD_relation}) means that the expected return ($\mu$) on
the time-$t$ fundamental value over the horizon $\tau$ is exactly offset by
the expected total future dividends re-invested at the risk-free rate to the
end of horizon $\tau$. This happens when the stock's dividend in each period
equals the expected return on the fundamental value (i.e., $x_{t}=\mu
F_{t}^{\mathbb{P}}$), which generates a martingale series (part (ii) of the proposition).
\end{remark}

Next, consider the $\mathbb{P}$-bubble process $B^{\mathbb{P}}$. Under the
rational bubble model (\ref{eq: recursive relation}), and given the
$\mathbb{P}$-bubble definition in (\ref{BP}), we see that $B^{\mathbb{P}}$ is
a non-negative process. A $\mathbb{P}$-bubble is said to occur at time $t$
when $B_{t}^{\mathbb{P}}$ is \emph{strictly} positive.\footnote{It can be
easily shown (using the recursive relation (\ref{EB})) that if a $\mathbb{P}%
$-bubble occurs at some time, it occurs at all time; see Proposition
\ref{propA_BP} and Remark \ref{rmkA_BP} in Section 2.3.}

The $\mathbb{P}$-bubble process displays certain dynamical properties. In
particular, if the stock earns a positive expected rate of return ($\mu>0$),
and a $\mathbb{P}$-bubble occurs at some time, then the expected size of the
$\mathbb{P}$-bubble will grow over time (although the statement does not say
anything about whether and when the $\mathbb{P}$-bubble will burst
probabilistically). This is stated below.\footnote{The submartingale property
depends on the assumption that $\mu>0$. This property will be lost if we relax
this assumption (e.g., if $\mu<0$ due to a negative risk adjustment associated
with the $\mathbb{P}$-bubble process); see lemma \ref{lemmaA_BP} in Section
2.3.}

\begin{lemma}
\label{lemma_BP}Suppose $\mu>0$. Then $B^{\mathbb{P}}$ is a $\mathbb{P}%
$-submartingale (i.e., $E_{t}^{\mathbb{P}}(B_{t+\tau}^{P})\geq B_{t}%
^{\mathbb{P}}$ ($\mathbb{P}$-a.s.) for all $\tau>0$). If additionally a
$\mathbb{P}$-bubble occurs at some time $t$, then $B^{\mathbb{P}}$ is a
\emph{strict} $\mathbb{P}$-submartingale (i.e., $E_{t}^{\mathbb{P}}(B_{t+\tau
}^{P})>B_{t}^{\mathbb{P}}$ ($\mathbb{P}$-a.s.) for all $\tau>0$).
\end{lemma}

\subsection{P-Martingale Deviation}

We define the $\mathbb{P}$-martingale deviation (at time $t$ over a time
horizon $\tau>0$) as
\[
\Pi_{t}^{\mathbb{P}}(\tau):=E_{t}^{\mathbb{P}}[S_{t+\tau}+D_{t:t+\tau}%
]-S_{t},
\]
where we recall that $D_{t:t+\tau}$ represents the time-$(t+\tau)$ value of
all dividends over $(t,t+\tau]$ re-invested at the risk-free
rate.\footnote{Assuming that the dividends are reinvested at the risk-free
rate is necessary to ensure no-arbitrage and maintain consistency in the
evaluation of expected returns under $\mathbb{P}$ and $\mathbb{Q}$ when we
calculate the equity risk premium; see Section 3. The dividends could
alternatively be reinvested in the stock, but the expressions would be more
complicated.}

The $\mathbb{P}$-martingale deviation consists of three parts: (1) the
expected change in the fundamental value; (2) the expected cumulative value of
all intermittent dividends re-invested at the risk-free rate; and (3) the
expected increment of the $\mathbb{P}$-bubble (if present) over the specified
time horizon. This is made precise in the following proposition.

\begin{proposition}
\label{prop_piP}The $\mathbb{P}$-martingale deviation $\Pi_{t}^{\mathbb{P}%
}(\tau)$ is given by
\begin{align}
\Pi_{t}^{\mathbb{P}}(\tau)  &  =\{E_{t}^{\mathbb{P}}(F_{t+\tau}^{\mathbb{P}%
})-F_{t}^{\mathbb{P}}\}+E_{t}^{\mathbb{P}}(D_{t:t+\tau})+\{E_{t}^{\mathbb{P}%
}(B_{t+\tau}^{\mathbb{P}})-B_{t}^{\mathbb{P}}\}\label{Pi_P}\\
&  =[(1+\mu)^{\tau}-1]F_{t}^{\mathbb{P}}-E_{t}^{\mathbb{P}}[(\tilde
{D}_{t:t+\tau}-D_{t:t+\tau})]+[(1+\mu)^{\tau}-1]B_{t}^{\mathbb{P}}.
\label{Pi_P2}%
\end{align}

\end{proposition}

In general, there is no guarantee that $\Pi_{t}^{\mathbb{P}}(\tau)$ is
nonnegative due to the negative term in (\ref{Pi_P2}); see Remark
\ref{rmrk_FP}. Exceptions occur in the following special cases.

\begin{corollary}
\label{coro_piP}$\Pi_{t}^{\mathbb{P}}(\tau)\geq0$ if any one of the following
conditions is true:

(i) $F^{\mathbb{P}}$ is a (sub)martingale under $\mathbb{P}$;

(ii) the stock pays no dividends;

(iii) $\mathbb{P}=\mathbb{Q}$ and $r_{t:t+1}^{f}$ is a constant.
\end{corollary}

The $\mathbb{P}$-martingale deviation is a useful metric for inferring the
existence of rational bubbles. Testing for a positive $\mathbb{P}$-martingale
deviation $\Pi_{t}^{\mathbb{P}}(\tau)>0$ is related to the rationale behind
some econometric methods testing for the presence of rational bubbles (e.g.,
Evans (1991), Phillips, Wu and Yu (2011), Phillips, Shi and Yu (2015a,b)).
These methods often require that the fundamental value process be a martingale
(i.e., case (i) in Corollary 1).\footnote{For example, if the dividend series
is assumed to follow a stationary ARMA model with independent innovations,
then for large horizon, the fundamental value process $F^{\mathbb{P}}$ is such
that $E_{t}^{\mathbb{P}}(F_{t+\tau}^{\mathbb{P}})$ grows linearly with $\tau$
for large $\tau$, suggesting that $\{F_{t}^{\mathbb{P}}\}$ contains a unit
root (Evans (1991)) and behaves like a unit-root martingale (c.f., Park and
Whang (2005)).} The observation that the dividend series is not explosive
(i.e., at most an I(1) process) and the price series is explosive is often
viewed as an evidence for the presence of $\mathbb{P}$-bubbles (e.g., Diba and
Grossman (1998), Evans (1991), Phillips, Wu and Yu (2011)).

The next corollary relates the existence of rational bubbles to the positivity
of the $\mathbb{P}$-martingale deviation.

\begin{corollary}
\label{coro_piP2}Suppose $\mu>0$ and that the fundamental value process
$F^{\mathbb{P}}$ is a martingale.

(i) If a $\mathbb{P}$-bubble exists, then $\Pi_{t}^{\mathbb{P}}(\tau)>0$ for
$t\geq0$ and $\tau>0$.

(ii) If additionally it is known that the stock pays no dividends over some
time window $[t,t+\tau]$, then a $\mathbb{P}$-bubble exists iff $\Pi
_{t}^{\mathbb{P}}(\tau)>0$.
\end{corollary}

\begin{example}
Suppose the price process $p$ of a stock (possibly including dividend
payments) follows an explosive AR(1) process given by $p_{t+1}=(1+\phi
)p_{t}+\varepsilon_{t+1}$, where $\phi>0$ and $\varepsilon_{t+1}$ is an iid
error. We deduce that, for $\tau\geq1$, $E_{t}^{\mathbb{P}}(p_{t+\tau
})=(1+\phi)^{\tau}p_{t}$, so that $\Pi_{t}^{\mathbb{P}}(\tau)=[(1+\phi)^{\tau
}-1]p_{t}>0$ (because $\phi>0$).
\end{example}

\subsection{Continuous-Time Extension with Stochastic Discount Factor}

Now we extend the classical rational bubble model by introducing a stochastic
discount factor and a continuous-time framework.\footnote{Hirano and Toda
(2024c) present the rational bubble model in a continuous-time framework
(Appendix B), but their focus is different and restrict themselves to the case
with no aggregate uncertainty; consequently, they do not treat the case where
the payoffs and the discount factor are stochastic.} The main difference from
the analysis in discrete time with a constant expected rate of change lies in
the presence of risk adjustment associated with the stock price and its
components. While the key expressions and results remain qualitatively similar
up to a risk-adjustment term, this leads to some subtle and yet important
changes in the dynamic properties of the $\mathbb{P}$-bubble process.

The important changes are summarized below.

\begin{itemize}
\item An equivalent way to ensure that fundamental value process
$F^{\mathbb{P}}$ is a $\mathbb{P}$-martingale is that the expected accumulated
dividends over a time horizon\ is equal to the risk-free return on
$F_{t}^{\mathbb{P}}$ over the horizon \emph{plus} expected compensation to
investors for bearing risks on the fundamental value process and accumulated
dividends over the same horizon. See Proposition \ref{propA_FP}.

\item In general, the $\mathbb{P}$-bubble process $B^{\mathbb{P}}$ is
\emph{not} a $\mathbb{P}$-submartingale because of the presence of the
risk-adjustment term involving the covariance of the stochastic discount
factor $m_{t+\tau}$ and the $\mathbb{P}$-bubble $B_{t+\tau}^{\mathbb{P}}$ at a
future time $t+\tau$; see Lemma \ref{lemmaA_BP}. In the special case where
$Cov_{t}^{\mathbb{P}}(m_{t+\tau},B_{t+\tau}^{\mathbb{P}})\leq0$ (e.g., when
the expected rate of return $\mu$ on stock is assumed to be a positive
constant as in lemma \ref{lemma_BP}), $B^{\mathbb{P}}$ is a $\mathbb{P}$-submartingale.

\item Risk adjustment plays a role in affecting the sign and magnitude of the
$\mathbb{P}$-martingale deviation $\Pi_{t}^{\mathbb{P}}(\tau)$; see
Proposition \ref{propA_piP}.
\end{itemize}

Let us introduce the continuous-time framework. The filtered probability space
remains to be $(\Omega,\mathcal{F}_{T},\mathbb{F},\mathbb{P})$. The market has
a riskless asset and a stock paying a continuous stream of dividends. Let
$S:=\{S_{t}\}_{t\in\lbrack0,T]}$\ denote the non-negative ex-dividend stock
price process. Let $X:=\{X_{t}\}_{t\in\lbrack0,T]}$ denote the non-negative
and non-decreasing cumulative dividend process. Both are assumed to be
$\mathbb{F}$-adapted processes.

Let $m_{t}$ denote the stochastic discount factor at time $t$. We assume that
$m_{t}>0$ for all $t$.

Let $R_{t:t+\tau}^{f}$ denote the gross risk-free rate over the period
$(t,t+\tau]$ on the riskless asset. It is assumed to be $\mathcal{F}_{t}%
$-measurable, so no randomness is involved in $R_{t:t+\tau}^{f}$ given the
information set $\mathcal{F}_{t}$\ as of time $t$. Because of this, we have
$1=E_{t}^{\mathbb{P}}(\frac{m_{t+\tau}}{m_{t}}R_{t:t+\tau}^{f})=\frac{1}%
{m_{t}}E_{t}^{\mathbb{P}}(m_{t+\tau})R_{t:t+\tau}^{f}$, so that $R_{t:t+\tau
}^{f}=\frac{m_{t}}{E_{t}^{\mathbb{P}}(m_{t+\tau})}$. We assume that
$R_{t:t+\tau}^{f}\geq1$ (equivalently $r_{t:t+\tau}^{f}\geq0$). This implies
$m_{t}\geq E_{t}^{\mathbb{P}}(m_{t+\tau})$, or $\{m_{t}\}_{t\in\lbrack0,T]}$
is a $\mathbb{P}$-supermartingale.

Given $s<t$, we define $D_{s:t}:=\int_{s}^{t}R_{u:t}^{f}dX_{u}$. It is the
time-$t$ future value of all dividends distributed over $(s,t]$ and reinvested
in the riskless asset up to time $t$. In addition, we define $\tilde{D}%
_{s:t}:=\int_{s}^{t}\frac{m_{u}}{m_{t}}dX_{u}$, the time-$t$ future value of
all dividends distributed over $(s,t]$ and reinvested in the risky asset up to
time $t$.

The price process $S$ satisfies the difference equation: for small
$\bigtriangleup>0$,
\[
m_{t}S_{t}=E_{t}^{\mathbb{P}}[m_{t+\bigtriangleup}(S_{t+\bigtriangleup
}+X_{t+\bigtriangleup}-X_{t})].
\]
The conditional expectation is taken under the statistical probability measure
$\mathbb{P}$.

Representing this in differential form, we have
\[
0=E_{t}^{\mathbb{P}}[d(m_{t}S_{t})+m_{t}dX_{t}].
\]
Integrating over $[t,T]$ and then taking conditional expectations yield
\[
m_{t}S_{t}=E_{t}^{\mathbb{P}}\left[  \int_{t}^{T}m_{u}dX_{u}\right]
+E_{t}^{\mathbb{P}}[m_{T}S_{T}].
\]

Dividing both sides by $m_{t}$ yields
\begin{equation}
S_{t}=\frac{1}{m_{t}}E_{t}^{\mathbb{P}}\left[  \int_{t}^{T}m_{u}dX_{u}\right]
+\frac{1}{m_{t}}E_{t}^{\mathbb{P}}[m_{T}S_{T}]. \label{S_decomp}%
\end{equation}
Now, letting $T\rightarrow\infty$, we obtain the stock price decomposition
\[
S_{t}=F_{t}^{\mathbb{P}}+B_{t}^{\mathbb{P}},
\]
where the first term on the right is the fundamental value, defined by
\[
F_{t}^{\mathbb{P}}:=\lim_{T\rightarrow\infty}F_{t}^{\mathbb{P}}(T)\text{,
\ where }F_{t}^{\mathbb{P}}(T):=\frac{1}{m_{t}}E_{t}^{\mathbb{P}}\left[
\int_{t}^{T}m_{u}dX_{u}\right]  ,
\]
and the second term is the $\mathbb{P}$-bubble, defined by
\[
B_{t}^{\mathbb{P}}:=\lim_{T\rightarrow\infty}B_{t}^{\mathbb{P}}(T)\text{,
\ where }B_{t}^{\mathbb{P}}(T):=\frac{1}{m_{t}}E_{t}^{\mathbb{P}}[m_{T}%
S_{T}].
\]
The decomposition is well defined as stated in the following lemma.

\begin{lemma}
\label{lemmaA_FPBP}The limits $\lim_{T\rightarrow\infty}E_{t}^{\mathbb{P}%
}\left[  \frac{1}{m_{t}}\int_{t}^{T}m_{u}dX_{u}\right]  $ and $\lim
_{T\rightarrow\infty}\frac{1}{m_{t}}E_{t}^{\mathbb{P}}[m_{T}S_{T}]$ exist.
\end{lemma}

\bigskip

The lemma below states that the time-$t$ fundamental value can be represented
as the present value of the fundamental value at $t+\tau$ plus the present
value of all intermittent dividends paid from time $t$ till $t+\tau$.

\begin{lemma}
\label{lemmaA_FP}The fundamental value is expressed as:
\[
F_{t}^{\mathbb{P}}=\frac{1}{m_{t}}E_{t}^{\mathbb{P}}\int_{t}^{t+\tau}%
m_{u}dX_{u}+\frac{1}{m_{t}}E_{t}^{\mathbb{P}}(m_{t+\tau}F_{t+\tau}%
^{\mathbb{P}}).
\]

\end{lemma}

Note that $F^{\mathbb{P}}$ may not be a $\mathbb{P}$-martingale. The dynamic
properties of $F^{\mathbb{P}}$ depend on how dividends are distributed over
time. A necessary and sufficient condition for $F^{\mathbb{P}}$ to be a
$\mathbb{P}$-martingale is provided in the following proposition.

\begin{proposition}
\label{propA_FP}We have
\begin{equation}
E_{t}^{\mathbb{P}}(F_{t+\tau}^{\mathbb{P}})-F_{t}^{\mathbb{P}}=(R_{t:t+\tau
}^{f}-1)F_{t}^{\mathbb{P}}-E_{t}^{\mathbb{P}}(\tilde{D}_{t:t+\tau
})-R_{t:t+\tau}^{f}Cov_{t}^{\mathbb{P}}(\frac{m_{t+\tau}}{m_{t}},F_{t+\tau
}^{\mathbb{P}}+\tilde{D}_{t:t+\tau});\label{EFFA}%
\end{equation}
in particular, $F^{\mathbb{P}}$ is a $\mathbb{P}$-martingale iff
\begin{equation}
(R_{t:t+\tau}^{f}-1)F_{t}^{\mathbb{P}}=E_{t}^{\mathbb{P}}(\tilde{D}_{t:t+\tau
})+R_{t:t+\tau}^{f}Cov_{t}^{\mathbb{P}}(\frac{m_{t+\tau}}{m_{t}},F_{t+\tau
}^{\mathbb{P}}+\tilde{D}_{t:t+\tau})\label{FD_relationA}%
\end{equation}
for all $\tau>0$.
\end{proposition}

In words, the fundamental value process is a $\mathbb{P}$-martingale if and
only if the risk-free return on $F_{t}^{\mathbb{P}}$ over period $(t,t+\tau]$
is used entirely to fund the dividends distributed over the same period (to be
reinvested in the stock), up to a risk adjustment associated with the
fundamental value $F_{t+\tau}^{\mathbb{P}}$ and the accumulated dividends
$\tilde{D}_{t:t+\tau}$ over the same period.

\begin{remark}
\label{rmrkA_FP}The fundamental value process can have very general dynamics.
In particular, the sign of $E_{t}^{\mathbb{P}}(F_{t+\tau}^{\mathbb{P}}%
)-F_{t}^{\mathbb{P}}$ can be positive or negative and time-varying, as there
are no restrictions on the sign of the risk adjustment term.
\end{remark}

\subsubsection*{}

Next, consider the $\mathbb{P}$-bubble process $B^{\mathbb{P}}$, defined by
\[
B_{t}^{\mathbb{P}}:=\lim_{T\rightarrow\infty}B_{t}^{\mathbb{P}}(T)\text{,
\ where }B_{t}^{\mathbb{P}}(T):=\frac{1}{m_{t}}E_{t}^{\mathbb{P}}[m_{T}%
S_{T}].
\]
The definition entails a number of properties $B^{\mathbb{P}}$ must satisfy.

\begin{proposition}
\label{propA_BP}The following statements are true.

(i) $B_{t}^{\mathbb{P}}\geq0$ for all $t$.

(ii) The discounted $\mathbb{P}$-bubble process $\{m_{t}B_{t}^{\mathbb{P}%
}\}_{0\leq t\leq T}$ is a $\mathbb{P}$-martingale.

(iii) $B_{t}^{\mathbb{P}}=0$ iff $B_{u}^{\mathbb{P}}=0$, $\mathbb{P}$-almost
surely, for all $u>t$.
\end{proposition}

\begin{remark}
\label{rmkA_BP}The last statement can be expressed equivalently as follows:
$B_{t}^{\mathbb{P}}>0$ iff $\mathbb{P}(B_{u}^{\mathbb{P}}>0)>0$ for all $u>t$.
In words, a $\mathbb{P}$-bubble occurs at time $t$ if and only if it occurs at
all future time $u>t$ with positive $\mathbb{P}$-probability. By repeated
application of the equivalence statement, we see that if a $\mathbb{P}$-bubble
occurs at some time, it occurs at all time.
\end{remark}

Let us analyze the dynamics of the $\mathbb{P}$-bubble process more closely.
It turns out that the discrepancy of $B^{\mathbb{P}}$ from a $\mathbb{P}%
$-martingale is determined by the risk-free return on the current $\mathbb{P}%
$-bubble plus a risk adjustment term on the $\mathbb{P}$-bubble process
involving the covariance of the $\mathbb{P}$-bubble and stochastic discount
factor in the future. This is stated in the following lemma.

\begin{lemma}
\label{lemmaA_BP}We have
\begin{equation}
E_{t}^{\mathbb{P}}(B_{t+\tau}^{\mathbb{P}})-B_{t}^{\mathbb{P}}=(R_{t:t+\tau
}^{f}-1)B_{t}^{\mathbb{P}}-R_{t:t+\tau}^{f}Cov_{t}^{\mathbb{P}}(\frac
{m_{t+\tau}}{m_{t}},B_{t+\tau}^{\mathbb{P}}).\label{EBBA}%
\end{equation}

\end{lemma}

\begin{remark}
The first term on the right of (\ref{EBBA}) is non-negative as $R_{t:t+\tau
}^{f}\geq1$. However, the sign of the risk adjustment term is indeterminate as
we do not have a priori knowledge about the sign of $Cov_{t}^{\mathbb{P}%
}(m_{t+\tau},B_{t+\tau}^{\mathbb{P}})$. Under the condition that
$Cov_{t}^{\mathbb{P}}(m_{t+\tau},B_{t+\tau}^{\mathbb{P}})\leq0$, the
$\mathbb{P}$-bubble process is a $\mathbb{P}$-submartingale; e.g., when there
is no uncertainty in the discount factor (hence no risk adjustment is necessary).
\end{remark}

Same as the discrete-time case, the $\mathbb{P}$-martingale deviation consists
of three parts: (1) the expected change in the fundamental value; (2) the
expected increment of the $\mathbb{P}$-bubble (if present) over the specified
time horizon; and (3) the expected cumulative value of all intermittent
dividends re-invested at the risk-free rate. This is made precise in the
following proposition.

\begin{proposition}
\label{propA_piP}The $\mathbb{P}$-martingale deviation $\Pi_{t}^{\mathbb{P}%
}(\tau)$ is given by
\begin{align}
\Pi_{t}^{\mathbb{P}}(\tau) &  =\{E_{t}^{\mathbb{P}}(F_{t+\tau}^{\mathbb{P}%
})-F_{t}^{\mathbb{P}}\}+\{E_{t}^{\mathbb{P}}(B_{t+\tau}^{\mathbb{P}}%
)-B_{t}^{\mathbb{P}}\}+E_{t}^{\mathbb{P}}(D_{t:t+\tau})\label{Pi_PA}\\
&  =(R_{t:t+\tau}^{f}-1)(F_{t}^{\mathbb{P}}+B_{t}^{\mathbb{P}})-E_{t}%
^{\mathbb{P}}[(\tilde{D}_{t:t+\tau}-D_{t:t+\tau})]\nonumber\\
&  -R_{t:t+\tau}^{f}Cov_{t}^{\mathbb{P}}(\frac{m_{t+\tau}}{m_{t}},F_{t+\tau
}^{\mathbb{P}}+B_{t+\tau}^{\mathbb{P}}+\tilde{D}_{t:t+\tau}).\label{Pi_P2A}%
\end{align}

\end{proposition}

In general, there is no guarantee that $\Pi_{t}^{\mathbb{P}}(\tau)$ is
non-negative. Exceptions occur under the following special cases. Compared
with Corollary \ref{coro_piP} in Section 2.2, the extra conditions on the
covariance between the stochastic discount factor and the stock price with
dividends are required to ensure that the risk adjustment is positive and does
not offset the risk-free growth rate on the stock price (first term on the
right of (\ref{Pi_P2})).

\begin{corollary}
\label{coroA_piP1}$\Pi_{t}^{\mathbb{P}}(\tau)\geq0$ if any one of the
following conditions is true:

(i) $F^{\mathbb{P}}$ and $B^{\mathbb{P}}$ are (sub)martingales under
$\mathbb{P}$;

(ii) the stock pays no dividends and $Cov_{t}^{\mathbb{P}}(m_{t+\tau
},S_{t+\tau})\leq0$;

(iii) $R_{s:t}^{f}=\frac{m_{s}}{m_{t}}$ for all $s<t$, and $Cov_{t}%
^{\mathbb{P}}(m_{t+\tau},S_{t+\tau}+\tilde{D}_{t:t+\tau})\leq0$.
\end{corollary}

The next corollary relates the existence of rational bubbles to the positivity
of the $\mathbb{P}$-martingale deviation. The extra assumption on
$Cov_{t}^{\mathbb{P}}(m_{t+\tau},B_{t+\tau}^{\mathbb{P}})$ restricts the risk
adjustment associated with the $\mathbb{P}$-bubble process, and is required to
accurately infer the presence of $\mathbb{P}$-bubbles using the metric
$\Pi_{t}^{\mathbb{P}}(\tau)$.

\begin{corollary}
\label{coroA_piP2}Suppose the fundamental value process $F^{\mathbb{P}}$ is a martingale.

(i) If a $\mathbb{P}$-bubble exists and $Cov_{t}^{\mathbb{P}}(m_{t+\tau
},B_{t+\tau}^{\mathbb{P}})\leq0$, then $\Pi_{t}^{\mathbb{P}}(\tau)>0$ for
$t\geq0$ and $\tau>0$.

(ii) If additionally the stock pays no dividends over the time window
$[t,t+\tau]$, and $Cov_{t}^{\mathbb{P}}(m_{t+\tau},B_{t+\tau}^{\mathbb{P}}%
)=0$, then a $\mathbb{P}$-bubble exists iff $\Pi_{t}^{\mathbb{P}}(\tau)>0$.
\end{corollary}

\section{A Local Martingale Bubble Model in Continuous Time}

Let us turn to the local martingale theory of bubbles (e.g., Lowenstein and
Willard (2000), Cox and Hobson (2005), Heston, Lowenstein and Willard (2007),
Jarrow, Protter and Shimbo (2010); see Protter (2013) for an overview). This
section describes the model (Section 3.1), introduces the $\mathbb{Q}%
$-martingale deviation measure used for empirical analysis (Section 3.2), and
shows that the local martingale bubble model includes the rational bubble
model as a special case (Section 3.3).

\subsection{Local Martingale Bubble Model}

In this model, time is set to be continuous.\footnote{The local martingale
theory of bubbles requires continuous time when characterizing the
$\mathbb{Q}$-bubble in the stock (defined as the difference between the spot
price and the fundamental value) as a strict local martingale process. In
discrete time, all local martingales reduce to martingales, and no
$\mathbb{Q}$-bubbles exist.} The model's horizon $T$ may be finite or
unbounded. The market for the stock and riskless asset is assumed to be
arbitrage-free in the sense that a no-free-lunch-with-vanishing-risk condition
is satisfied. This guarantees the existence of a risk-neutral probability
measure $\mathbb{Q}$ (more precisely, an equivalent local martingale measure);
see Delbaen and Schachermayer (1994).

The definition of the risk-neutral measure $\mathbb{Q}$ is such that the
process $\{(R_{0:t}^{f})^{-1}(S_{t}+D_{0:t})\}_{t\geq0}$, stock price plus
cumulative dividends discounted at the risk-free rate, is a local martingale
under $\mathbb{Q}$.

We can decompose the normalized ex-dividend price process plus accumulated
dividends as follows:
\begin{equation}
S_{t}+D_{0:t}=F_{t}^{\mathbb{Q}}(T)+B_{t}^{\mathbb{Q}}(T), \label{Qdecomp}%
\end{equation}
where
\[
F_{t}^{\mathbb{Q}}(T):=E_{t}^{\mathbb{Q}}[(R_{t:T}^{f})^{-1}(S_{T}+D_{0:T})],
\]
and
\[
B_{t}^{\mathbb{Q}}(T):=S_{t}+D_{0:t}-F_{t}^{\mathbb{Q}}(T)
\]
with $B_{T}^{\mathbb{Q}}(T)=0$.

Here, $F_{t}^{\mathbb{Q}}(T)$ corresponds to the stock's fundamental value,
which is the $\mathbb{Q}$-expectation of the discounted payoff of the stock
plus accumulated dividends. The residual $B_{t}^{\mathbb{Q}}(T)$ represents
the deviation of the stock price from its fundamental value. It is the type-3
bubble in the local martingale theory (Jarrow, Protter and Shimbo (2010)). We
call $\{F_{t}^{\mathbb{Q}}(T)\}_{0\leq t\leq T}$ the fundamental value process
under $\mathbb{Q}$, and $\{B_{t}^{\mathbb{Q}}(T)\}_{0\leq t\leq T}$ the
$\mathbb{Q}$-bubble process.

Note that the fundamental value process satisfies the following dynamic
relation, which directly follows from the definition. It states that the
time-$t$ fundamental value is determined by the present value of the
time-$(t+\tau)$ future fundamental value.

\begin{lemma}
\label{lemma_FQ}$F_{t}^{\mathbb{Q}}(T)=E_{t}^{\mathbb{Q}}[(R_{t:t+\tau}%
^{f})^{-1}F_{t+\tau}^{\mathbb{Q}}(T)]$.
\end{lemma}

The next proposition characterizes the dynamics of the discounted stock price
plus accumulated dividend process and the discounted fundamental value
process. Note that there is an adjustment for risk when computing the
fundamental value of stock under $\mathbb{Q}$, so that the fundamental value
discounted at the risk-free rate is a martingale process.

\begin{proposition}
\label{prop_SDFD}(i) $\{(R_{0:t}^{f})^{-1}(S_{t}+D_{0:t})\}_{t\in\lbrack0,T]}$
is a $\mathbb{Q}$-supermartingale.

(ii) $\{(R_{0:t}^{f})^{-1}F_{t}^{\mathbb{Q}}(T)\}_{t\in\lbrack0,T]}$ is a
$\mathbb{Q}$-martingale.

(iii) $\{(R_{0:t}^{f})^{-1}B_{t}^{\mathbb{Q}}(T)\}_{t\in\lbrack0,T]}$ is a
$\mathbb{Q}$-supermartingale.
\end{proposition}

As a corollary, we have the following inequality, which follows directly from
the proposition.

\begin{corollary}
\label{coro_SSD}The following inequalities hold true:

(i) $S_{t}\geq E_{t}^{\mathbb{Q}}[(R_{t:t+\tau}^{f})^{-1}(S_{t+\tau
}+D_{t:t+\tau})].$

(ii) $B_{t}^{\mathbb{Q}}(T)\geq E_{t}^{\mathbb{Q}}[(R_{t:t+\tau}^{f}%
)^{-1}B_{t+\tau}^{\mathbb{Q}}(T)].$
\end{corollary}

The $\mathbb{Q}$-bubble process is always non-negative ($B_{t}^{\mathbb{Q}%
}(T)\geq0$ for all $t\in\lbrack0,T]$) with a zero terminal value
($B_{T}^{\mathbb{Q}}(T)=0$). A $\mathbb{Q}$-bubble is said to occur at time
$t$ when $B_{t}^{\mathbb{Q}}(T)$ is \emph{strictly} positive. If a
$\mathbb{Q}$-bubble occurs at some time $t$, then the discounted $\mathbb{Q}%
$-bubble process $\{(R_{0:t}^{f})^{-1}B_{t}^{\mathbb{Q}}(T)\}_{t\in
\lbrack0,T]}$ is a \emph{strict} supermartingale under $\mathbb{Q}$.

\subsection{Q-Martingale Deviation}

We define the $\mathbb{Q}$-martingale deviation (at time $t$, over a time
horizon $\tau$, where $0<\tau\leq T-t$) as follows:
\begin{equation}
\Pi_{t}^{\mathbb{Q}}(\tau):=S_{t}-E_{t}^{\mathbb{Q}}[(R_{t:t+\tau}^{f}%
)^{-1}(S_{t+\tau}+D_{t:t+\tau})]. \label{Pi_Q}%
\end{equation}
Note that $\Pi_{t}^{\mathbb{Q}}(\tau)\geq0$ by Corollary \ref{coro_SSD}(i).

The following proposition shows that the $\mathbb{Q}$-martingale deviation
captures the expected decrease in discounted $\mathbb{Q}$-bubble, thus
justifying its use for $\mathbb{Q}$-bubble analysis in empirical work.

\begin{proposition}
\label{prop_PiQ}We have
\[
\Pi_{t}^{\mathbb{Q}}(\tau)=B_{t}^{\mathbb{Q}}(T)-E_{t}^{\mathbb{Q}%
}[(R_{t:t+\tau}^{f})^{-1}B_{t+\tau}^{\mathbb{Q}}(T)].
\]

\end{proposition}

\subsection{Relationship with Classical Rational Bubble Model}

This section shows that the local martingale bubble model includes the
classical rational bubble model as a special case. The key result is stated in
the following proposition. Note that the local martingale bubble model reduces
to the rational bubble model (\ref{Pdecomp}) if: (i) the model's horizon
approaches infinity, and (ii) we exclude the $\mathbb{Q}$-bubble.

\begin{proposition}
\label{prop_PQ}Under the local martingale bubble model given in (\ref{Qdecomp}%
), the stock price can be decomposed as
\[
S_{t}=F_{t}^{\mathbb{P}}(T)+B_{t}^{\mathbb{P}}(T)+B_{t}^{\mathbb{Q}}(T),
\]
where $F_{t}^{\mathbb{P}}(T)$ is as defined in (\ref{FP}), $B_{t}^{\mathbb{P}%
}(T)$ is as defined in (\ref{BP}), and $B_{t}^{\mathbb{Q}}(T)$ is as defined
in (\ref{Qdecomp}). This reduces to the stock price decomposition
(\ref{Pdecomp}) under the rational bubble model by letting $T\rightarrow
\infty$ and assuming that $\lim_{T\rightarrow\infty}B_{t}^{\mathbb{Q}}(T)=0$
for all $t$.
\end{proposition}

\section{Equity Risk Premium}

This section joins the two different bubble theories in order to obtain a
characterization of the equity risk premium involving the $\mathbb{P}$- and
$\mathbb{Q}$-bubbles. This characterization is new to the literature and forms
the basis for our subsequent empirical estimation.

Let $S_{t}$ denote the ex-dividend stock price. Let $r_{t:t+\tau}%
:=\frac{S_{t+\tau}-S_{t}+D_{t:t+\tau}}{S_{t}}$ denote the simple return on the
stock plus dividends over the period $(t,t+\tau]$. Let $r_{t:t+\tau}%
^{f}:=R_{t:t+\tau}^{f}-1$ denote the net return on the riskless asset over the
same period. Due to their riskless nature, $R_{t:t+\tau}^{f}$ and
$r_{t:t+\tau}^{f}$ are deterministic conditional on the information set at
time $t$, although they are stochastic unconditionally.

The equity risk premium (ERP) is defined to be the difference in conditional
mean return evaluated under $\mathbb{P}$ and $\mathbb{Q}$:
\[
ERP_{t}(\tau)=E_{t}^{\mathbb{P}}(r_{t:t+\tau})-E_{t}^{\mathbb{Q}}(r_{t:t+\tau
}).
\]

It is possible to decompose ERP and express it in terms of the two martingale
deviations. This is formally given in the following proposition.

\begin{proposition}
\label{prop_ERP}We have
\[
ERP_{t}(\tau)=\underset{ERP_{t}^{\mathbb{P}}(\tau)}{\underbrace{\frac{\Pi
_{t}^{\mathbb{P}}(\tau)}{S_{t}}}}+\underset{ERP_{t}^{\mathbb{Q}}(\tau
)}{\underbrace{R_{t:t+\tau}^{f}\frac{\Pi_{t}^{\mathbb{Q}}(\tau)}{S_{t}%
}-r_{t:t+\tau}^{f}}},
\]
where $ERP_{t}^{\mathbb{P}}(\tau):=E_{t}^{\mathbb{P}}(r_{t:t+\tau})$ and
$ERP_{t}^{\mathbb{Q}}(\tau):=-E_{t}^{\mathbb{Q}}(r_{t:t+\tau})$ denote the
$\mathbb{P}$- and $\mathbb{Q}$-components of ERP, respectively.
\end{proposition}

In the standard no-arbitrage case where the stock price is a \emph{martingale}
under the \emph{equivalent martingale measure} $\mathbb{Q}$, the $\mathbb{Q}%
$-bubble is ruled out, so that $B_{t}^{\mathbb{Q}}(T)\equiv0$, $\Pi
_{t}^{\mathbb{Q}}(\tau)\equiv0$, and $E_{t}^{\mathbb{Q}}(r_{t:t+\tau
})=r_{t:t+\tau}^{f}$. The ERP thus reduces to the classical formula
\begin{equation}
ERP0_{t}(\tau)=E_{t}^{\mathbb{P}}(r_{t:t+\tau})-r_{t:t+\tau}^{f}=\frac{\Pi
_{t}^{\mathbb{P}}(\tau)}{S_{t}}-r_{t:t+\tau}^{f}. \label{ERP0}%
\end{equation}

The rational bubble theory underestimates the true ERP when a $\mathbb{Q}%
$-bubble is present. This is stated in the following corollary.

\begin{corollary}
\label{coro_ERP}We have%
\[
ERP_{t}(\tau)=ERP0_{t}(\tau)+R_{t:t+\tau}^{f}\frac{\Pi_{t}^{\mathbb{Q}}(\tau
)}{S_{t}}\geq ERP0_{t}(\tau),
\]
and the inequality is strict if $\mathbb{Q}$-bubble exists.
\end{corollary}

Our result generalizes the existing literature on ERP estimation using market
data. Under the standard no-arbitrage condition and a negative correlation
condition, Martin (2017) derives the following lower bound for $ERP0_{t}%
(\tau)$:%
\[
ERP0_{t}(\tau)\geq R_{t:t+\tau}^{f}\cdot\tau\cdot SVIX_{t\rightarrow t+\tau
}^{2},
\]
where $\tau\cdot SVIX_{t\rightarrow t+\tau}^{2}=Var_{t}^{\mathbb{Q}}\left(
\frac{R_{t:t+\tau}}{R_{t:t+\tau}^{f}}\right)  $ measures the risk-neutral
return variance, which can be estimated using option prices (and the spot and
futures prices of the underlying) at time $t$ (see equations (12)-(13) in
Martin (2017)). Under the stronger NFLVR condition, the lower bound needs to
be revised upward as suggested by Corollary \ref{coro_ERP}:%
\[
ERP_{t}(\tau)\geq R_{t:t+\tau}^{f}\cdot\tau\cdot SVIX_{t\rightarrow t+\tau
}^{2}+R_{t:t+\tau}^{f}\frac{\Pi_{t}^{\mathbb{Q}}(\tau)}{S_{t}}.
\]
The upward adjustment $R_{t:t+\tau}^{f}\frac{\Pi_{t}^{\mathbb{Q}}(\tau)}%
{S_{t}}$ represents the necessary compensation for the anticipated price drop
due to the presence of a $\mathbb{Q}$-bubble (in such case the price process
is a $\mathbb{Q}$-supermartingale).\footnote{Recall that the $\mathbb{Q}%
$-martingale deviation $\Pi_{t}^{\mathbb{Q}}(\tau)$ captures the downward
adjustment in the first $\mathbb{Q}$-moment of the stock price (plus
dividends) due to the $\mathbb{Q}$-bubble. If a $\mathbb{Q}$-bubble exists,
investors anticipate a negative expected stock return under $\mathbb{Q}$ (with
a present value equal to $-\frac{\Pi_{t}^{\mathbb{Q}}(\tau)}{S_{t}}$ as of
time $t$). To cover this anticipated loss, they will require risk compensation
equal to $R_{t:t+\tau}^{f}\frac{\Pi_{t}^{\mathbb{Q}}(\tau)}{S_{t}}$ as of time
$t+\tau$.} The second term involving $\frac{\Pi_{t}^{\mathbb{Q}}(\tau)}{S_{t}%
}$ captures the negative risk-neutral expected return (a first-moment
property); whereas the first term involving $SVIX_{t\rightarrow t+\tau}^{2}$
measures the risk-neutral return variance (a second-moment property). Both
terms contribute to the ERP in a complementary way.

\section{Concluding Remark}

This paper relates the two types of bubbles defined in the literature. One is
based on the classical rational bubble model, and the second is based on the
local martingale bubble model, which is shown to be more general. We link both
types of bubbles to an equity's risk premium via a decomposition of this risk
premium into components related to both bubbles. The analysis is important for
understanding the contribution to the equity's risk premium and is potentially
useful for empirical asset pricing.

\section*{Appendix: Technical Proofs}

\subsection*{A1 Proof of Lemma \ref{lemma_FP}}

We start from the definition of $F_{t}^{\mathbb{P}}$ and rewrite it as
follows:
\begin{align*}
F_{t}^{\mathbb{P}} &  =\sum_{j=1}^{\infty}\frac{1}{(1+\mu)^{j}}E_{t}%
^{\mathbb{P}}(x_{t+j})\\
&  =\sum_{j=1}^{\tau}\frac{1}{(1+\mu)^{j}}E_{t}^{\mathbb{P}}(x_{t+j}%
)+\sum_{j=\tau+1}^{\infty}\frac{1}{(1+\mu)^{j}}E_{t}^{\mathbb{P}}(x_{t+j})\\
&  =\frac{1}{(1+\mu)^{\tau}}\sum_{j=1}^{\tau}E_{t}^{\mathbb{P}}[(1+\mu
)^{\tau-j}x_{t+j}]+\frac{1}{(1+\mu)^{\tau}}\sum_{j=1}^{\infty}\frac{1}%
{(1+\mu)^{j}}E_{t}^{\mathbb{P}}[E_{t+\tau}^{\mathbb{P}}(x_{t+\tau+j})]\\
&  =\frac{1}{(1+\mu)^{\tau}}E_{t}^{\mathbb{P}}(\tilde{D}_{t:t+\tau})+\frac
{1}{(1+\mu)^{\tau}}E_{t}^{\mathbb{P}}(F_{t+\tau}^{\mathbb{P}}),
\end{align*}
as claimed.

\subsection*{A2 Proof of Proposition \ref{prop_FP}}

\noindent(i) From lemma \ref{lemma_FP}, we have (after mutiplying both sides
by $(1+\mu)^{\tau}$)
\[
(1+\mu)^{\tau}F_{t}^{\mathbb{P}}=E_{t}^{\mathbb{P}}(F_{t+\tau}^{\mathbb{P}%
}+\tilde{D}_{t:t+\tau}),
\]
which implies
\[
E_{t}^{\mathbb{P}}(F_{t+\tau}^{\mathbb{P}})-F_{t}^{\mathbb{P}}=[(1+\mu)^{\tau
}-1]F_{t}^{\mathbb{P}}-E_{t}^{\mathbb{P}}(\tilde{D}_{t:t+\tau}).
\]
This equals $0$ iff $[(1+\mu)^{\tau}-1]F_{t}^{\mathbb{P}}=E_{t}^{\mathbb{P}%
}(\tilde{D}_{t:t+\tau})$ for all $\tau>0$, as stated.

\noindent(ii) If $E_{t}^{\mathbb{P}}(x_{t+j})=x_{t}=\mu F_{t}^{\mathbb{P}}$,
then expression (\ref{EFF}) simplifies to
\begin{align*}
E_{t}^{\mathbb{P}}(F_{t+\tau}^{\mathbb{P}})-F_{t}^{\mathbb{P}}  &
=[(1+\mu)^{\tau}-1]F_{t}^{\mathbb{P}}-\sum_{j=1}^{\tau}(1+\mu)^{\tau-j}%
E_{t}^{\mathbb{P}}(x_{t+j})\\
&  =[(1+\mu)^{\tau}-1]F_{t}^{\mathbb{P}}-\sum_{j=1}^{\tau}(1+\mu)^{\tau
-j}x_{t}\\
&  =[(1+\mu)^{\tau}-1]F_{t}^{\mathbb{P}}-(1+\mu)^{\tau}\frac{\frac{1}{1+\mu
}\left[  1-\frac{1}{(1+\mu)^{\tau}}\right]  }{1-\frac{1}{1+\mu}}x_{t}\\
&  =[(1+\mu)^{\tau}-1]F_{t}^{\mathbb{P}}-[(1+\mu)^{\tau}-1]\frac{1}{\mu}%
x_{t}\\
&  =0.
\end{align*}

\subsection*{A3 Proof of Lemma \ref{lemma_BP}}

From expression (\ref{EB}), we have $E_{t}^{\mathbb{P}}(B_{t+1}^{\mathbb{P}%
})=(1+\mu)B_{t}^{\mathbb{P}}\geq B_{t}^{\mathbb{P}}$, and so $B^{\mathbb{P}}$
is a $\mathbb{P}$-submartingale. If $B_{t}^{\mathbb{P}}>0$ for some time $t$,
then $B_{t}^{\mathbb{P}}>0$ for all $t$. It follows that $E_{t}^{\mathbb{P}%
}(B_{t+1}^{\mathbb{P}})=(1+\mu)B_{t}^{\mathbb{P}}>B_{t}^{\mathbb{P}}$, and so
$B^{\mathbb{P}}$ is a \emph{strict} $\mathbb{P}$-submartingale.

\subsection*{A4 \ Proof of Proposition \ref{prop_piP}}

Using the decomposition (\ref{Pdecomp}), we obtain
\begin{align*}
\Pi_{t}^{\mathbb{P}}(\tau)  &  =E_{t}^{\mathbb{P}}(S_{t+\tau}+D_{t:t+\tau
})-S_{t}\\
&  =E_{t}^{\mathbb{P}}(F_{t+\tau}^{\mathbb{P}}+B_{t+\tau}^{\mathbb{P}}%
)-(F_{t}^{\mathbb{P}}+B_{t}^{\mathbb{P}})+E_{t}^{\mathbb{P}}(D_{t:t+\tau})\\
&  =\{E_{t}^{\mathbb{P}}(F_{t+\tau}^{\mathbb{P}})-F_{t}^{\mathbb{P}}%
\}+E_{t}^{\mathbb{P}}(D_{t:t+\tau})+\{E_{t}^{\mathbb{P}}(B_{t+\tau
}^{\mathbb{P}})-B_{t}^{\mathbb{P}}\}\\
&  =[(1+\mu)^{\tau}-1]F_{t}^{\mathbb{P}}-E_{t}^{\mathbb{P}}[\tilde
{D}_{t:t+\tau}]+E_{t}^{\mathbb{P}}(D_{t:t+\tau})+[(1+\mu)^{\tau}%
-1]B_{t}^{\mathbb{P}},
\end{align*}
where the last step is obtained from (\ref{EFF}) in Proposition \ref{prop_FP}%
(i) and by repeated use of (\ref{EB}).

\subsection*{A5 \ Proof of Corollary \ref{coro_piP}}

\noindent(i) If $F^{\mathbb{P}}$ is a $\mathbb{P}$-(sub)martingale, then the
first curly bracket in (\ref{Pi_P}) is nonnegative, and hence $\Pi
_{t}^{\mathbb{P}}(\tau)=E_{t}^{\mathbb{P}}(D_{t:t+\tau})+\{E_{t}^{\mathbb{P}%
}(B_{t+\tau}^{\mathbb{P}})-B_{t}^{\mathbb{P}}\}\geq0$ (because dividends are
non-negative and $B^{\mathbb{P}}$ is a submartingale).

\noindent(ii) If the stock pays no dividends, then the first two curly
brackets in (\ref{Pi_P}) are equal to zero, and hence $\Pi_{t}^{\mathbb{P}%
}(\tau)=E_{t}^{\mathbb{P}}(B_{t+\tau}^{\mathbb{P}})-B_{t}^{\mathbb{P}}\geq0$.

\noindent(iii) In the special case where $\mathbb{P}=\mathbb{Q}$, the expected
rate of return on stock plus dividend is the same as the constant risk-free
rate (i.e., $1+\mu=e^{r_{t:t+1}^{f}}$ for all $t$), and hence $E_{t}%
^{\mathbb{P}}[(\tilde{D}_{t:t+\tau}-D_{t:t+\tau})]=0$ (i.e., the middle term
in (\ref{Pi_P2}) vanishes). It follows that $\Pi_{t}^{\mathbb{P}}%
(\tau)=[(1+\mu)^{\tau}-1](F_{t}^{\mathbb{P}}+B_{t}^{\mathbb{P}})\geq0$.

\subsection*{A6 \ Proof of Corollary \ref{coro_piP2}}

\noindent(i) From the assumptions, we have $F_{t}^{\mathbb{P}}=E_{t}%
^{\mathbb{P}}(F_{t+\tau}^{\mathbb{P}})$, and hence $\Pi_{t}^{\mathbb{P}}%
(\tau)=E_{t}^{\mathbb{P}}(D_{t:t+\tau})+\{E_{t}^{\mathbb{P}}(B_{t+\tau
}^{\mathbb{P}})-B_{t}^{\mathbb{P}}\}$ by equation (\ref{Pi_P}). If
$\mathbb{P}$-bubble exists, then $B^{\mathbb{P}}$ is a strict $\mathbb{P}%
$-submartingale, i.e., $E_{t}^{\mathbb{P}}(B_{t+\tau}^{\mathbb{P}}%
)>B_{t}^{\mathbb{P}}$ for $\tau>0$. This implies $\Pi_{t}^{\mathbb{P}}%
(\tau)>0$ for $\tau>0$.

\noindent(ii) We now have $D_{t:t+\tau}=0$. It follows from equation
(\ref{Pi_P}) that $\Pi_{t}^{\mathbb{P}}(\tau)=E_{t}^{\mathbb{P}}(B_{t+\tau
}^{\mathbb{P}})-B_{t}^{\mathbb{P}}$. The equivalence is seen immediately.

\subsection*{A7 \ Proof of Lemma \ref{lemmaA_FPBP}}

For fixed $t$, note that $f_{t}(T):=\frac{1}{m_{t}}\int_{t}^{T}m_{u}dX_{u}$ is
non-negative and increasing with $T$ (because $X$ is a non-decreasing process
and $m_{t}>0$ for all $t$). It follows that the first limit $\lim
_{T\rightarrow\infty}E_{t}^{\mathbb{P}}\left[  \frac{1}{m_{t}}\int_{t}%
^{T}m_{u}dX_{u}\right]  =\lim_{T\rightarrow\infty}E_{t}^{\mathbb{P}}\left[
f_{t}(T)\right]  =E_{t}^{\mathbb{P}}\left[  f_{t}(\infty)\right]
=E_{t}^{\mathbb{P}}\left[  \frac{1}{m_{t}}\int_{t}^{\infty}m_{u}dX_{u}\right]
$ exists by the monotone convergence theorem. The second limit $\lim
_{T\rightarrow\infty}\frac{1}{m_{t}}E_{t}^{\mathbb{P}}[m_{T}S_{T}]$ is equal
to $S_{t}-\lim_{T\rightarrow\infty}\frac{1}{m_{t}}E_{t}^{\mathbb{P}}\left[
\int_{t}^{T}m_{u}dX_{u}\right]  $ by (\ref{S_decomp}) and is well defined.

\subsection*{A8 Proof of Lemma \ref{lemmaA_FP}}

We have
\begin{align*}
m_{t}F_{t}^{\mathbb{P}} &  =E_{t}^{\mathbb{P}}\int_{t}^{\infty}m_{u}dX_{u}\\
&  =E_{t}^{\mathbb{P}}\int_{t}^{t+\tau}m_{u}dX_{u}+E_{t}^{\mathbb{P}}%
E_{t+\tau}^{\mathbb{P}}\int_{t+\tau}^{\infty}m_{u}dX_{u}\\
&  =E_{t}^{\mathbb{P}}\int_{t}^{t+\tau}m_{u}dX_{u}+E_{t}^{\mathbb{P}%
}(m_{t+\tau}F_{t+\tau}^{\mathbb{P}}).
\end{align*}
Dividing both sides by $m_{t}$ immediately yields the result.

\subsection*{A9 Proof of Proposition \ref{propA_FP}}

Using lemma \ref{lemmaA_FP}, we have
\begin{align*}
m_{t}F_{t}^{\mathbb{P}} &  =E_{t}^{\mathbb{P}}\int_{t}^{t+\tau}m_{u}%
dX_{u}+E_{t}^{\mathbb{P}}(m_{t+\tau}F_{t+\tau}^{\mathbb{P}})\\
&  =E_{t}^{\mathbb{P}}\int_{t}^{t+\tau}m_{u}dX_{u}+E_{t}^{\mathbb{P}%
}(m_{t+\tau})E_{t}^{\mathbb{P}}(F_{t+\tau}^{\mathbb{P}})+Cov_{t}^{\mathbb{P}%
}(m_{t+\tau},F_{t+\tau}^{\mathbb{P}}).
\end{align*}
Solving for $E_{t}^{\mathbb{P}}(F_{t+\tau}^{\mathbb{P}})$ gives
\begin{align}
E_{t}^{\mathbb{P}}(F_{t+\tau}^{\mathbb{P}}) &  =\frac{m_{t}}{E_{t}%
^{\mathbb{P}}(m_{t+\tau})}F_{t}^{\mathbb{P}}-\frac{1}{E_{t}^{\mathbb{P}%
}(m_{t+\tau})}E_{t}^{\mathbb{P}}\int_{t}^{t+\tau}m_{u}dX_{u}-\frac{1}%
{E_{t}^{\mathbb{P}}(m_{t+\tau})}Cov_{t}^{\mathbb{P}}(m_{t+\tau},F_{t+\tau
}^{\mathbb{P}})\nonumber\\
&  =\frac{m_{t}}{E_{t}^{\mathbb{P}}(m_{t+\tau})}F_{t}^{\mathbb{P}}-\frac
{1}{E_{t}^{\mathbb{P}}(m_{t+\tau})}E_{t}^{\mathbb{P}}\int_{t}^{t+\tau}%
m_{u}dX_{u}-\frac{m_{t}}{E_{t}^{\mathbb{P}}(m_{t+\tau})}Cov_{t}^{\mathbb{P}%
}(\frac{m_{t+\tau}}{m_{t}},F_{t+\tau}^{\mathbb{P}})\nonumber\\
&  =R_{t:t+\tau}^{f}F_{t}^{\mathbb{P}}-\frac{1}{E_{t}^{\mathbb{P}}(m_{t+\tau
})}E_{t}^{\mathbb{P}}\int_{t}^{t+\tau}m_{u}dX_{u}-R_{t:t+\tau}^{f}%
Cov_{t}^{\mathbb{P}}(\frac{m_{t+\tau}}{m_{t}},F_{t+\tau}^{\mathbb{P}%
}),\label{EFA}%
\end{align}
where the last line uses the definition of $R_{t:t+\tau}^{f}$. The second term
on the right is interpreted as the risk-adjusted version of $\tilde
{D}_{t:t+\tau}$. Indeed, it can be expressed as
\begin{align*}
&  \frac{1}{E_{t}^{\mathbb{P}}(m_{t+\tau})}E_{t}^{\mathbb{P}}\int_{t}^{t+\tau
}m_{u}dX_{u}\\
&  =E_{t}^{\mathbb{P}}\left[  \frac{m_{t+\tau}}{E_{t}^{\mathbb{P}}(m_{t+\tau
})}\int_{t}^{t+\tau}\frac{m_{u}}{m_{t+\tau}}dX_{u}\right]  \\
&  =E_{t}^{\mathbb{P}}\left[  \frac{m_{t+\tau}}{E_{t}^{\mathbb{P}}(m_{t+\tau
})}\tilde{D}_{t:t+\tau}\right]  \\
&  =E_{t}^{\mathbb{P}}\left[  \frac{m_{t+\tau}}{E_{t}^{\mathbb{P}}(m_{t+\tau
})}\right]  E_{t}^{\mathbb{P}}(\tilde{D}_{t:t+\tau})+Cov_{t}^{\mathbb{P}%
}\left[  \frac{m_{t+\tau}}{E_{t}^{\mathbb{P}}(m_{t+\tau})},\tilde{D}%
_{t:t+\tau}\right]  \\
&  =E_{t}^{\mathbb{P}}(\tilde{D}_{t:t+\tau})+Cov_{t}^{\mathbb{P}}\left[
\frac{m_{t+\tau}}{E_{t}^{\mathbb{P}}(m_{t+\tau})},\tilde{D}_{t:t+\tau}\right]
.
\end{align*}
Substituting into (\ref{EFA}) and subtracting $F_{t}^{\mathbb{P}}$ from both
sides, we obtain
\[
E_{t}^{\mathbb{P}}(F_{t+\tau}^{\mathbb{P}})-F_{t}^{\mathbb{P}}=(R_{t:t+\tau
}^{f}-1)F_{t}^{\mathbb{P}}-E_{t}^{\mathbb{P}}(\tilde{D}_{t:t+\tau
})-R_{t:t+\tau}^{f}Cov_{t}^{\mathbb{P}}(\frac{m_{t+\tau}}{m_{t}},F_{t+\tau
}^{\mathbb{P}}+\tilde{D}_{t:t+\tau}).
\]

\subsection*{A10 Proof of Proposition \ref{propA_BP}}

\noindent(i) This is immediate from the definition of $B_{t}^{\mathbb{P}}$ and
the assumptions that $m_{t},m_{T}>0$ and $S_{T}\geq0$.

\noindent(ii) To see this, note that, for all $\tau>0$
\begin{align}
m_{t}B_{t}^{\mathbb{P}} &  =\lim_{T\rightarrow\infty}E_{t}^{\mathbb{P}}%
(m_{T}S_{T})\nonumber\\
&  =\lim_{T\rightarrow\infty}E_{t}^{\mathbb{P}}E_{t+\tau}^{\mathbb{P}}%
(m_{T}S_{T}).\nonumber\\
&  =E_{t}^{\mathbb{P}}(m_{t+\tau}B_{t+\tau}^{\mathbb{P}}),\label{EmB}%
\end{align}
where the last step follows from the monotone convergence theorem (for any
given $t+\tau$, the function $g_{t+\tau}(T)=E_{t+\tau}^{\mathbb{P}}(m_{T}%
S_{T})=m_{t+\tau}S_{t+\tau}-E_{t+\tau}^{\mathbb{P}}(\int_{t+\tau}^{T}%
m_{u}dX_{u})$ decreases with $T$, so $\lim_{T\rightarrow\infty}E_{t}%
^{\mathbb{P}}[g_{t+\tau}(T)]=E_{t}^{\mathbb{P}}[\lim_{T\rightarrow\infty
}g_{t+\tau}(T)]=E_{t}^{\mathbb{P}}(m_{t+\tau}B_{t+\tau}^{\mathbb{P}})$ by the
definition of $\mathbb{P}$-bubble).

\noindent(iii) Applying (\ref{EmB}) with $u=t+\tau$ yields
\begin{equation}
B_{t}^{\mathbb{P}}=E_{t}^{\mathbb{P}}(\frac{m_{u}}{m_{t}}B_{u}^{\mathbb{P}%
}).\label{EmmB}%
\end{equation}

\textquotedblleft If\textquotedblright\ part: Because $B_{u}^{\mathbb{P}}=0$
$\mathbb{P}$-almost surely and $\frac{m_{u}}{m_{t}}$ is finite, this implies
that $B_{t}^{\mathbb{P}}=E_{t}^{\mathbb{P}}(\frac{m_{u}}{m_{t}}B_{u}%
^{\mathbb{P}})=0$.

\textquotedblleft Only if\textquotedblright\ part: Because $\frac{m_{u}}%
{m_{t}}B_{u}^{\mathbb{P}}\geq0$, $\mathbb{P}$-almost surely, it follows from
$0=B_{t}^{\mathbb{P}}=E_{t}^{\mathbb{P}}(\frac{m_{u}}{m_{t}}B_{u}^{\mathbb{P}%
})$ that, $\mathbb{P}$-almost surely, $\frac{m_{u}}{m_{t}}B_{u}^{\mathbb{P}%
}=0$, and hence $B_{u}^{\mathbb{P}}=0$ for all $u>t$.

\subsection*{A11 Proof of Lemma \ref{lemmaA_BP}}

Recall that
\[
m_{t}B_{t}^{\mathbb{P}}=E_{t}^{\mathbb{P}}(m_{t+\tau}B_{t+\tau}^{\mathbb{P}%
})=E_{t}^{\mathbb{P}}(m_{t+\tau})E_{t}^{\mathbb{P}}(B_{t+\tau}^{\mathbb{P}%
})+Cov_{t}^{\mathbb{P}}(m_{t+\tau},B_{t+\tau}^{\mathbb{P}}),
\]
which implies that
\begin{align*}
E_{t}^{\mathbb{P}}(B_{t+\tau}^{\mathbb{P}}) &  =\frac{m_{t}}{E_{t}%
^{\mathbb{P}}(m_{t+\tau})}B_{t}^{\mathbb{P}}-\frac{m_{t}}{E_{t}^{\mathbb{P}%
}(m_{t+\tau})}Cov_{t}^{\mathbb{P}}(\frac{m_{t+\tau}}{m_{t}},B_{t+\tau
}^{\mathbb{P}})\\
&  =R_{t:t+\tau}^{f}\left[  B_{t}^{\mathbb{P}}-Cov_{t}^{\mathbb{P}}%
(\frac{m_{t+\tau}}{m_{t}},B_{t+\tau}^{\mathbb{P}})\right]  ,
\end{align*}
or
\[
E_{t}^{\mathbb{P}}(B_{t+\tau}^{\mathbb{P}})-B_{t}^{\mathbb{P}}=(R_{t:t+\tau
}^{f}-1)B_{t}^{\mathbb{P}}-R_{t:t+\tau}^{f}Cov_{t}^{\mathbb{P}}(\frac
{m_{t+\tau}}{m_{t}},B_{t+\tau}^{\mathbb{P}}).
\]

\subsection*{A12 Proof of Proposition \ref{propA_piP}}

The first equality is by the decomposition $S=F^{\mathbb{P}}+B^{\mathbb{P}}$.
The second equality follows from Proposition \ref{propA_FP} and lemma
\ref{lemmaA_BP}.

\subsection*{A13 Proof of Corollary \ref{coroA_piP1}}

\noindent(i) follows directly from (\ref{Pi_P}).

\noindent(ii) follows from (\ref{Pi_P2}), noting that $\tilde{D}_{t:t+\tau
}=D_{t:t+\tau}=0$.

\noindent(iii) follows from (\ref{Pi_P2}), noting that the conditions imply
$\tilde{D}_{t:t+\tau}=D_{t:t+\tau}$.

\subsection*{A14 Proof of Corollary \ref{coroA_piP2}}

\noindent(i) From the assumptions, we have $F_{t}^{\mathbb{P}}=E_{t}%
^{\mathbb{P}}(F_{t+\tau}^{\mathbb{P}})$, and hence $\Pi_{t}^{\mathbb{P}}%
(\tau)=E_{t}^{\mathbb{P}}(D_{t:t+\tau})+\{E_{t}^{\mathbb{P}}(B_{t+\tau
}^{\mathbb{P}})-B_{t}^{\mathbb{P}}\}$ by equation (\ref{Pi_PA}). If a
$\mathbb{P}$-bubble exists and $Cov_{t}^{\mathbb{P}}(m_{t+\tau},B_{t+\tau
}^{\mathbb{P}})\leq0$, then lemma \ref{lemmaA_BP} implies that $B^{\mathbb{P}%
}$ is a strict submartingale, i.e., $E_{t}^{\mathbb{P}}(B_{t+\tau}%
^{\mathbb{P}})>B_{t}^{\mathbb{P}}$ for $\tau>0$. This implies $\Pi
_{t}^{\mathbb{P}}(\tau)>0$ for $\tau>0$.

\noindent(ii) We now have $D_{t:t+\tau}=0$ and $Cov_{t}^{\mathbb{P}}%
(m_{t+\tau},B_{t+\tau}^{\mathbb{P}})=0$. It follows from equation
(\ref{Pi_PA}) and lemma \ref{lemmaA_BP} that $\Pi_{t}^{\mathbb{P}}(\tau
)=E_{t}^{\mathbb{P}}(B_{t+\tau}^{\mathbb{P}})-B_{t}^{\mathbb{P}}=(R_{t:t+\tau
}^{f}-1)B_{t}^{\mathbb{P}}$. The equivalence is seen immediately.

\subsection*{A15 Proof of Lemma \ref{lemma_FQ}}

Starting from the right side,
\begin{align*}
RHS  &  =E_{t}^{\mathbb{Q}}[(R_{t:t+\tau}^{f})^{-1}F_{t+\tau}^{\mathbb{Q}%
}(T)]\\
&  =E_{t}^{\mathbb{Q}}\{(R_{t:t+\tau}^{f})^{-1}E_{t+\tau}^{\mathbb{Q}%
}[(R_{t+\tau:T}^{f})^{-1}(S_{T}+D_{0:T})]\}\\
&  =E_{t}^{\mathbb{Q}}\{E_{t+\tau}^{\mathbb{Q}}[(R_{t:t+\tau}^{f}%
)^{-1}(R_{t+\tau:T}^{f})^{-1}(S_{T}+D_{0:T})]\}\\
&  =E_{t}^{\mathbb{Q}}\{(R_{t:T}^{f})^{-1}(S_{T}+D_{0:T})\}\\
&  =F_{t}^{\mathbb{Q}}(T)\\
&  =LHS,
\end{align*}
where step 2 uses the definition of $F_{t+\tau}^{\mathbb{Q}}(T)$ and step 4
uses the tower rule.

\subsubsection*{}

\subsection*{A16 Proof of Proposition \ref{prop_SDFD}}

\noindent(i) By the non-negativity of stock price and accumulated dividends,
and the fact that a non-negative local martingale is a supermartingale.

\noindent(ii) This follows directly from lemma \ref{lemma_FQ} because it
implies $(R_{0:t}^{f})^{-1}F_{t}^{\mathbb{Q}}(T)=E_{t}^{\mathbb{Q}%
}[(R_{0:t+\tau}^{f})^{-1}F_{t+\tau}^{\mathbb{Q}}(T)]$.

\noindent(iii) This follows from (i)-(ii) and the decomposition in
(\ref{Qdecomp}).

\subsection*{A17 Proof of Corollary \ref{coro_SSD}}

\noindent(i) From Proposition \ref{prop_SDFD}(i), we have
\[
(R_{0:t}^{f})^{-1}(S_{t}+D_{0:t})\geq E_{t}^{\mathbb{Q}}[(R_{0:t+\tau}%
^{f})^{-1}(S_{t+\tau}+D_{0:t+\tau})],
\]
or
\[
S_{t}+D_{0:t}\geq E_{t}^{\mathbb{Q}}[(R_{t:t+\tau}^{f})^{-1}(S_{t+\tau
}+D_{0:t+\tau})].
\]
Given relation (\ref{D}) that
\[
D_{0:t+\tau}=D_{0:t}R_{t:t+\tau}^{f}+D_{t:t+\tau},
\]
substitution gives
\[
(R_{t:t+\tau}^{f})^{-1}D_{0:t+\tau}=D_{0:t}+(R_{t:t+\tau}^{f})^{-1}%
D_{t:t+\tau}.
\]
Applying to the inequality yields
\[
S_{t}\geq E_{t}^{\mathbb{Q}}[(R_{t:t+\tau}^{f})^{-1}(S_{t+\tau}+D_{t:t+\tau
})].
\]

\noindent(ii) By Proposition \ref{prop_SDFD}(i), lemma \ref{lemma_FQ}, and a
simple subtraction, the result follows immediately.

\subsection*{A18\ Proof of Proposition \ref{prop_PiQ}}

Given the definition of $\Pi_{t}^{\mathbb{Q}}(\tau)$, we have
\begin{align*}
\Pi_{t}^{\mathbb{Q}}(\tau)  &  =S_{t}-E_{t}^{\mathbb{Q}}[(R_{t:t+\tau}%
^{f})^{-1}(S_{t+\tau}+D_{t:t+\tau})]\\
&  =S_{t}+D_{0:t}-E_{t}^{\mathbb{Q}}[(R_{t:t+\tau}^{f})^{-1}(S_{t+\tau
}+D_{0:t+\tau})]\\
&  =F_{t}^{\mathbb{Q}}(T)+B_{t}^{\mathbb{Q}}(T)-E_{t}^{\mathbb{Q}%
}[(R_{t:t+\tau}^{f})^{-1}(F_{t+\tau}^{\mathbb{Q}}(T)+B_{t+\tau}^{\mathbb{Q}%
}(T))]\\
&  =B_{t}^{\mathbb{Q}}(T)-E_{t}^{\mathbb{Q}}[(R_{t:t+\tau}^{f})^{-1}B_{t+\tau
}^{\mathbb{Q}}(T)],
\end{align*}
where the second step uses relation (\ref{D}), the third step makes use of the
decomposition in (\ref{Qdecomp}), and the last step applies lemma
\ref{lemma_FQ}.

\subsection*{A19 Proof of Proposition \ref{prop_PQ}}

Let $\Lambda_{t}:=\left.  \frac{d\mathbb{Q}}{d\mathbb{P}}\right\vert
_{\mathcal{F}_{t}}$ denote the Radon-Nikodym derivative. The stochastic
discount factor is given by $m_{t}=(R_{0:t}^{f})^{-1}\Lambda_{t}$.

Using relation (\ref{D}) on dividends, we can express the fundamental value
as:
\begin{align*}
F_{t}^{\mathbb{Q}}(T)  &  =E_{t}^{\mathbb{Q}}[(R_{t:T}^{f})^{-1}(D_{0:T}%
+S_{T})]\\
&  =D_{0:t}+E_{t}^{\mathbb{Q}}[(R_{t:T}^{f})^{-1}(D_{t:T}+S_{T})].
\end{align*}
After applying a change of measure, we obtain
\begin{align*}
F_{t}^{\mathbb{Q}}(T)-D_{0:t}  &  =E_{t}^{\mathbb{Q}}[(R_{t:T}^{f}%
)^{-1}(D_{t:T}+S_{T})]\\
&  =E_{t}^{\mathbb{Q}}\left[  \int_{t}^{T}(R_{t:u}^{f})^{-1}dX_{u}\right]
+E_{t}^{\mathbb{Q}}[(R_{t:T}^{f})^{-1}S_{T}]\\
&  =\Lambda_{t}^{-1}E_{t}^{\mathbb{P}}\left[  \int_{t}^{T}(R_{t:u}^{f}%
)^{-1}\Lambda_{u}dX_{u}\right]  +\Lambda_{t}^{-1}E_{t}^{\mathbb{P}}%
[(R_{t:T}^{f})^{-1}\Lambda_{T}S_{T}]\\
&  =m_{t}^{-1}E_{t}^{\mathbb{P}}\left[  \int_{t}^{T}m_{u}dX_{u}\right]
+m_{t}^{-1}E_{t}^{\mathbb{P}}[m_{T}S_{T}]\\
&  =F_{t}^{\mathbb{P}}(T)+B_{t}^{\mathbb{P}}(T).
\end{align*}
Substituting into (\ref{Qdecomp}) yields
\begin{align*}
S_{t}  &  =F_{t}^{\mathbb{Q}}(T)-D_{0:t}+B_{t}^{\mathbb{Q}}(T)\\
&  =F_{t}^{\mathbb{P}}(T)+B_{t}^{\mathbb{P}}(T)+B_{t}^{\mathbb{Q}}(T).
\end{align*}

\subsection*{A20 Proof of Proposition \ref{prop_ERP}}

Recall that the ERP is given by%
\[
ERP_{t}(\tau)=E_{t}^{\mathbb{P}}(r_{t:t+\tau})-E_{t}^{\mathbb{Q}}(r_{t:t+\tau
}).
\]
The first term on the right is given by
\[
E_{t}^{\mathbb{P}}(r_{t:t+\tau})=\frac{E_{t}^{\mathbb{P}}(S_{t+\tau
}+D_{t:t+\tau})-S_{t}}{S_{t}}=\frac{\Pi_{t}^{\mathbb{P}}(\tau)}{S_{t}}.
\]
To simplify the second term, we decompose the return into excess return and
risk-free rate:
\begin{align*}
E_{t}^{\mathbb{Q}}(r_{t:t+\tau})  &  =E_{t}^{\mathbb{Q}}[(1+r_{t:t+\tau
})(R_{t:t+\tau}^{f})^{-1}+(1+r_{t:t+\tau})(1-(R_{t:t+\tau}^{f})^{-1})-1]\\
&  =\frac{E_{t}^{\mathbb{Q}}[S_{t}(1+r_{t:t+\tau})(R_{t:t+\tau}^{f}%
)^{-1}-S_{t}]}{S_{t}}+E_{t}^{\mathbb{Q}}\left[  (1+r_{t:t+\tau})\left(
1-\frac{1}{1+r_{t:t+\tau}^{f}}\right)  \right] \\
&  =\frac{E_{t}^{\mathbb{Q}}[(S_{t+\tau}+D_{t:t+\tau})(R_{t:t+\tau}^{f}%
)^{-1}]-S_{t}}{S_{t}}+E_{t}^{\mathbb{Q}}\left[  (1+r_{t:t+\tau})\frac
{r_{t:t+\tau}^{f}}{1+r_{t:t+\tau}^{f}}\right] \\
&  =-\frac{\Pi_{t}^{\mathbb{Q}}(\tau)}{S_{t}}+r_{t:t+\tau}^{f}\frac
{1+E_{t}^{\mathbb{Q}}(r_{t:t+\tau})}{1+r_{t:t+\tau}^{f}},
\end{align*}
where the last step follows by recalling the definition of $\Pi_{t}%
^{\mathbb{Q}}(\tau)$ and noting that $r_{t:t+\tau}^{f}$ is deterministic given
the information set at time $t$.

From the last step, we can collect like terms and solve for $E_{t}%
^{\mathbb{Q}}(r_{t:t+\tau})$:%
\[
E_{t}^{\mathbb{Q}}(r_{t:t+\tau})=-R_{t:t+\tau}^{f}\frac{\Pi_{t}^{\mathbb{Q}%
}(\tau)}{S_{t}}+r_{t:t+\tau}^{f}.
\]

The equity risk premium is thus given by
\[
ERP_{t}(\tau)=E_{t}^{\mathbb{P}}(r_{t:t+\tau})-E_{t}^{\mathbb{Q}}(r_{t:t+\tau
})=\underset{ERP_{t}^{\mathbb{P}}(\tau)}{\underbrace{\frac{\Pi_{t}%
^{\mathbb{P}}(\tau)}{S_{t}}}}+\underset{ERP_{t}^{\mathbb{Q}}(\tau
)}{\underbrace{R_{t:t+\tau}^{f}\frac{\Pi_{t}^{\mathbb{Q}}(\tau)}{S_{t}%
}-r_{t:t+\tau}^{f}}}.
\]

\subsection*{A21 Proof of Corollary \ref{coro_ERP}}

The ERP can be decomposed as follows%
\begin{align*}
ERP_{t}(\tau)  &  =E_{t}^{\mathbb{P}}(r_{t:t+\tau})-E_{t}^{\mathbb{Q}%
}(r_{t:t+\tau})\\
&  =[E_{t}^{\mathbb{P}}(r_{t:t+\tau})-r_{t:t+\tau}^{f}]+[r_{t:t+\tau}%
^{f}-E_{t}^{\mathbb{Q}}(r_{t:t+\tau})].
\end{align*}
The first term is $E_{t}^{\mathbb{P}}(r_{t:t+\tau})-r_{t:t+\tau}^{f}%
=ERP0_{t}(\tau)$. The second term can be further simplified, using the result
in Proposition \ref{prop_ERP}.
\begin{align*}
r_{t:t+\tau}^{f}-E_{t}^{\mathbb{Q}}(r_{t:t+\tau})  &  =r_{t:t+\tau}^{f}%
+\frac{\Pi_{t}^{\mathbb{Q}}(\tau)}{S_{t}}-r_{t:t+\tau}^{f}\frac{1+E_{t}%
^{\mathbb{Q}}(r_{t:t+\tau})}{1+r_{t:t+\tau}^{f}}\\
&  =\frac{\Pi_{t}^{\mathbb{Q}}(\tau)}{S_{t}}+r_{t:t+\tau}^{f}\left(
1-\frac{1+E_{t}^{\mathbb{Q}}(r_{t:t+\tau})}{1+r_{t:t+\tau}^{f}}\right) \\
&  =\frac{\Pi_{t}^{\mathbb{Q}}(\tau)}{S_{t}}+\frac{r_{t:t+\tau}^{f}%
}{1+r_{t:t+\tau}^{f}}[r_{t:t+\tau}^{f}-E_{t}^{\mathbb{Q}}(r_{t:t+\tau})].
\end{align*}
Collecting like terms, we obtain%
\[
\left(  1-\frac{r_{t:t+\tau}^{f}}{1+r_{t:t+\tau}^{f}}\right)  [r_{t:t+\tau
}^{f}-E_{t}^{\mathbb{Q}}(r_{t:t+\tau})]=\frac{\Pi_{t}^{\mathbb{Q}}(\tau
)}{S_{t}},
\]
thus yielding the solution%
\[
r_{t:t+\tau}^{f}-E_{t}^{\mathbb{Q}}(r_{t:t+\tau})=[1+r_{t:t+\tau}^{f}%
]\frac{\Pi_{t}^{\mathbb{Q}}(\tau)}{S_{t}}=R_{t:t+\tau}^{f}\frac{\Pi
_{t}^{\mathbb{Q}}(\tau)}{S_{t}}.
\]
Note that $r_{t:t+\tau}^{f}-E_{t}^{\mathbb{Q}}(r_{t:t+\tau})\geq0$ by the
nonnegativity of $R_{t:t+\tau}^{f}$, $\Pi_{t}^{\mathbb{Q}}(\tau)$ and $S_{t}$.

When a $\mathbb{Q}$-bubble exists, $\{(R_{t:t+\tau}^{f})^{-1}(S_{t+\tau
}+D_{t:t+\tau})\}_{\tau>0}$ is a strict supermartingale, so that $\Pi
_{t}^{\mathbb{Q}}(\tau)>0$, and the weak inequality becomes strict. The proof
is now complete.

\end{document}